\documentclass[aps,preprint]{revtex4}
\usepackage{amssymb}
\usepackage{amsmath}
\usepackage{graphicx}
\usepackage{subfigure}

\begin{document}

\title{Lovelock Thin-Shell Wormholes}
\author{M. H. Dehghani $^{1,2}$ \footnote{email address:
mhd@shirazu.ac.ir} and M. R. Mehdizadeh $^{1}$}
\affiliation{$^1$Physics Department and Biruni Observatory,
College of Sciences, Shiraz University, Shiraz 71454, Iran\\
$^2$Research Institute for Astrophysics and Astronomy of Maragha
(RIAAM), Maragha, Iran}

\begin{abstract}
We construct the
asymptotically flat charged thin-shell wormholes of Lovelock
gravity in seven dimensions by cut-and-paste technique, and apply
the generalized junction conditions in order to calculate the
energy-momentum tensor of these wormholes on the shell. We find
that for negative second order and positive third order Lovelock coefficients, there are thin-shell
wormholes that respect the weak energy condition.
In this case, the amount of normal matter decreases as the third order Lovelock coefficient increases.
For positive second and third order Lovelock
coefficients, the weak energy condition is violated  and the amount of exotic matter decreases as the
charge increases. Finally, we perform a linear stability analysis
against a symmetry preserving perturbation, and find that the
wormholes are stable provided the derivative of surface pressure density with
respect to surface energy density is negative and the throat radius is chosen suitable.
\end{abstract}

\maketitle

\section{Introduction}

Traversable wormholes are throat like geometrical structures which connect
two separate and distinct regions of spacetimes and have no horizon or
singularity \cite{MT}. It is known that the traversable wormholes in
Einstein gravity possess a stress-energy tensor that violates the standard
energy conditions and therefore they are supported by exotic matter (see,
e.g., \cite{MTY}). There are two main areas in wormhole research which
attracted many authors.

The first one is to try avoiding, as much as
possible, the violation of the standard energy conditions. The
existence of traversable wormholes that are supported by arbitrarily small
quantities of exotic matter \cite{viskardad} or supported by matter not
violating the energy conditions \cite{Mart Rich,Deh1} have been
investigated. One of the most interesting kinds of traversable wormholes is
the thin-shell wormholes which are constructed by the cut-and-paste
technique used for the first time in relation to wormholes in Refs. \cite{Vis,VisP}.
This is due to the fact that energy is concentrated on the throat of
thin-shell wormholes, and therefore the construction of these wormholes
needs less exotic matter. Thin-shell wormholes have been investigated by
many authors \cite{Thin}.

The second main research area is the stability analysis of thin-shell
wormholes against a symmetry preserving perturbation. This can be done by considering a linearized stability analysis
around the static wormhole solutions. The stability analysis will tell whether a
static wormhole solution holds under small perturbations or not. The stability
analysis of four-dimensional thin-shell wormholes of Einstein gravity with
spherical symmetry in vacuum has been done in \cite{VisP} and in the presence
of cosmological constant in \cite{Lobo}. The generalization of these
stability analysis to the case of higher-dimensional wormholes can be found
in \cite{Lem}. Also for stability analysis with dilaton, axion, phantom and
other types of matter, and in cylindrical symmetry see \cite{Axion}.

For thin-shell wormholes of Einstein gravity, the weak energy condition (WEC)
is violated. Several attempts have been made to somehow overcome
this problem. Some authors resort to the alternative
theories of gravity. In this context, the thin-shell wormholes of dilaton gravity supported
by Chaplygin gas have been investigated in \cite{Dil}, of Brans-Dicke theory
in \cite{Brans} and of Gauss-Bonnet gravity in \cite{Gauss}.

Recently, the third order Lovelock gravity have attracted more attention
\cite{Deh2}. This is due to the fact that it contains more free parameters,
and therefore third order Lovelock gravity may be dual to a
more extended class of field theories which one can study with
holography \cite{Lg3}. Here, we like to add the third order term of Lovelock theory \cite{Lov} to
the gravitational field equations, and investigate the effects of it on the
energy conditions and stability of a thin-shell wormhole solution.

In order to do these, one needs the generalized junction conditions in Lovelock gravity.
The junction conditions in Einstein gravity is known as
Darmois-Israel junction conditions \cite{Isr}. Also the generalized junction
conditions in Gauss-Bonnet gravity \cite{Dav} and Lovelock gravity \cite{Grav} have been introduced.
In this paper, we use the generalized junction conditions in order to
investigate the exoticity of matter on the throat and to perform a linear
stability analysis of the thin-shell wormholes of Lovelock gravity.

The paper is organized as follows. In Sec. \ref{Sol}, we review the
asymptotically flat static charged solutions of third order Lovelock gravity
in seven dimensions. Section \ref{Junc} is devoted to the generalized junction conditions in
Lovelock gravity. We construct the thin-shell wormholes in Sec. \ref{Thin}.
In Sec. \ref{Exot}, we investigate the energy conditions on shell for these
wormhole solutions and we consider the exoticity of matter for different
values of the parameters of Lovelock gravity. Finally, we perform the
stability analysis of the wormhole solutions in Sec. \ref{Stab} . We finish our paper with some
concluding remarks.

\section{Spherically symmetric geometry\label{Sol}}

Here, we review the asymptotically flat charged static solutions of third order
Lovelock gravity. The action of third order Lovelock gravity in the presence
of electromagnetic field may be written as
\begin{equation}
I=\int d^{n+1}x\sqrt{-g}\left( \mathcal{L}_{1}+\alpha _{2}\mathcal{L}%
_{2}+\alpha _{3}\mathcal{L}_{3}-F_{\mu \nu }F^{\mu \nu }\right)  \label{Act1}
\end{equation}
where $\alpha _{2}$ and $\alpha _{3}$ are second (Gauss-Bonnet) and third
order Lovelock coefficients, $\mathcal{L}_{1}=R$ is just the
Einstein-Hilbert Lagrangian, $\mathcal{L}_{2}=R_{\mu \nu \gamma \delta
}R^{\mu \nu \gamma \delta }-4R_{\mu \nu }R^{\mu \nu }+R^{2}$ is the
Gauss-Bonnet Lagrangian, and
\begin{eqnarray}
\mathcal{L}_{3} &=&2R^{\mu \nu \sigma \kappa }R_{\sigma \kappa \rho \tau }R_{%
\phantom{\rho \tau }{\mu \nu }}^{\rho \tau }+8R_{\phantom{\mu \nu}{\sigma
\rho}}^{\mu \nu }R_{\phantom {\sigma \kappa} {\nu \tau}}^{\sigma \kappa }R_{%
\phantom{\rho \tau}{ \mu \kappa}}^{\rho \tau }+24R^{\mu \nu \sigma \kappa
}R_{\sigma \kappa \nu \rho }R_{\phantom{\rho}{\mu}}^{\rho }  \nonumber \\
&&+3RR^{\mu \nu \sigma \kappa }R_{\sigma \kappa \mu \nu }+24R^{\mu \nu
\sigma \kappa }R_{\sigma \mu }R_{\kappa \nu }+16R^{\mu \nu }R_{\nu \sigma
}R_{\phantom{\sigma}{\mu}}^{\sigma }-12RR^{\mu \nu }R_{\mu \nu }+R^{3}
\label{L3}
\end{eqnarray}
is the third order Lovelock Lagrangian. In Eq. (\ref{Act1}) $F_{\mu \nu
}=\partial _{\mu }A_{\nu }-\partial _{\nu }A_{\mu }$ is electromagnetic
tensor field and $A_{\mu }$ is the vector potential.

The seven-dimensional static solution of action (\ref{Act1}) may be written as
\begin{equation}
ds^{2}=-f(r)dt^{2}+\frac{dr^{2}}{f(r)}+r^{2}d\Omega _{5}^{2},  \label{met}
\end{equation}
where $d\Omega _{5}^{2}$ is the metric of a $5$-sphere, the gauge field is
\begin{equation*}
A=\sqrt{\frac{5}{8}}\frac{q}{r^{4}}dt,
\end{equation*}
and the metric function satisfies the following equation
\begin{equation}
24\alpha _{3}(1-f)^{3}+12\alpha _{2}(1-f)^{2}r^{2}+(1-f)r^{4}-m+%
\frac{q^{2}}{r^{4}}=0.  \label{Eq}
\end{equation}
The solution of Eq. (\ref{Eq}) is \cite{Deh2005}
\begin{equation}
f(r)=1+\frac{\alpha _{2}}{6\alpha _{3}}r^{2}+\frac{1}{12\alpha _{3}}%
\left( \zeta ^{1/3}-\frac{2(\alpha _{3}-2\alpha
_{2}^{2})r^{4}}{\zeta ^{1/3}}\right) ,  \label{Fgen}
\end{equation}
where $\zeta $ and $\xi $\ are
\begin{eqnarray*}
\zeta &=&\xi +\sqrt{\xi ^{2}+8(\alpha _{3}-2\alpha _{2}^{2})^{3}r^{12}}, \\
\xi &=&2\alpha _{2}(4\alpha _{2}^{2}-3\alpha _{3})r^{6}-36\alpha
_{3}^{2}(m-q^{2}r^{-4}).
\end{eqnarray*}
The above metric presents an
asymptotically flat black hole with two
horizons provided $q<$ $q_{ext}$, an extreme black hole with one horizon if $%
q=$ $q_{ext}$ and a naked singularity otherwise, where $q_{ext}$ is
\begin{eqnarray}
q_{ext} &=&\left\{\frac{2187}{2}\alpha _{2}^{4}-(3\alpha
_{3}-\frac{m}{8})(\eta
^{2}+243\alpha _{2}^{2})-\frac{3}{2}\alpha _{2}\eta ^{3}\right\}^{1/2},  \nonumber \\
\eta &=&\left(2m+81\alpha _{2}^{2}-48\alpha _{3}\right)^{1/2}. \label{qext7}
\end{eqnarray}

The metric function for the special case $\alpha _{3}=2{\alpha _{2}}%
^{2}=\alpha ^{2}/72$ reduces to
\begin{equation}
f(r)=1+{\frac{{r}^{2}}{\alpha }}\left\{ 1-\left( {1+\frac{3\alpha m}{r^{6}}-%
\frac{3\alpha {q}^{2}}{r^{10}}}\right) ^{1/3}\right\} .
\label{F7sp}
\end{equation}
In studying wormholes, the matter is outside the horizon $r>r_{h}$, where $%
r_{h}$ is the largest real root of $f(r)=0$.

\section{Junction Conditions in Lovelock Gravity \label{Junc}}

The action of Lovelock gravity with well-defined variational principle in $%
n+1$ dimension may be written as \cite{DBS}
\begin{eqnarray}
I_{G} &=&\kappa \int d^{n+1}x\sqrt{-g}\sum_{p=0}^{[n/2]}\alpha _{p}\mathcal{L}%
_{p}  \nonumber \\
&&-2\kappa \int_{\Sigma }d^{n}x\sqrt{-\gamma }\sum_{p=0}^{[n/2]}%
\sum_{s=0}^{p-1}\frac{(-1)^{p-s}p\alpha _{p}}{2^{s}(2p-2s-1)}\mathcal{H}%
^{(p)}  \label{Lov1}
\end{eqnarray}
where $[z]$ denotes the integer part of $z$, $\alpha _{p}$ is Lovelock
coefficient, $\mathcal{L}_{p}$ is the Euler density of a $2p$-dimensional
manifold
\begin{equation}
\mathcal{L}_{p}=\frac{1}{2^{p}}\delta _{\rho _{1}\sigma _{1}\cdots \rho
_{p}\sigma _{p}}^{\mu _{1}\nu _{1}\cdots \mu _{p}\nu _{p}}R_{\mu _{1}\nu
_{1}}^{\phantom{\mu_1\nu_1}{\rho_1\sigma_1}}\cdots R_{\mu _{p}\nu _{p}}^{%
\phantom{\mu_k \nu_k}{\rho_p \sigma_p}},  \label{Lov2}
\end{equation}
and
\begin{equation}
\mathcal{H}^{(p)}=\delta _{\,[b_{1}\ldots b_{2p-1}]}^{[a_{1}\ldots
a_{2p-1}]}R_{\phantom{b_1b_1}{a_1a_2}}^{b_{1}b_{2}}\cdots R_{%
\phantom{b_{2s}}{a_{2s-1}a_{2s}}}^{b_{2s-1}b_{2s}}\Theta _{a1}^{b1}\cdots
\Theta _{a_{2p-1}}^{b_{2p-1}}  \label{Hp}
\end{equation}
In the above equations $\delta _{\rho _{1}\sigma _{1}\cdots \rho _{p}\sigma
p}^{\mu _{1}\nu _{1}\cdots \mu p\nu _{p}}$ is the generalized totally
antisymmetric Kronicker delta and $\gamma _{ab}$ and $\Theta _{ab}$ are the
induced metric and extrinsic curvature of the timelike hypersurface $\Sigma $.

Varying the action (\ref{Lov1}) with respect to the induced metric $\gamma
_{ab}$ gives \cite{DBS,Grav}
\begin{equation}
\mathcal{T}_{b}^{a}=-2\kappa \sum_{p=0}^{n}\sum_{s=0}^{p-1}\frac{%
4^{p-s}p!\alpha
_{p}}{2^{p+1}s!(2p-2s-1)!!}\mathcal{H}_{\phantom{i}b}^{(p,s)a}
\label{Stres1}
\end{equation}
where $\mathcal{H}_{\phantom{i}b}^{(p,s)a}$ is
\begin{equation}
\mathcal{H}_{\phantom{i}b}^{(p,s)a}=\delta _{\,[b_{1}\ldots
b_{2p-1}b]}^{[a_{1}\ldots a_{2p-1}a]}R_{\phantom{b_1b_2}{a_1a_2}%
}^{b_{1}b_{2}}\cdots R_{\phantom{b_{2s-1}b_{2s}}{a_{2s-1}a_{2s}}%
}^{b_{2s-1}b_{2s}}\Theta _{a_{2s+1}}^{b_{2s+1}}\cdots \Theta
_{a_{2p-1}}^{b_{2p-1}}.  \label{Hps}
\end{equation}
In Eqs. (\ref{Hp}) and (\ref{Hps}) $R_{abcd}$ is the boundary
components of the Riemann tensor of the Manifold $\mathcal{M}$
which is related to the intrinsic curvature of the hypersurface
$\Sigma $, $\hat{R}_{a b c d}(\gamma )$, through the
Gauss-Codazzi equation as
\begin{equation*}
R_{abcd}=\hat{R}_{abcd}(\gamma)+\Theta _{ac}\Theta _{bd}-\Theta _{ad}\Theta _{bc}
\end{equation*}

Now considering the manifold $\mathcal{M}$ as two manifold $\mathcal{M}_{+}$
and $\mathcal{M}_{-}$ separated by hypersurface $\Sigma $, and denoting its
two sides by $\Sigma _{\pm }$, then the generalized junction conditions
can be written as
\begin{equation}
\mathcal{S}^{ab}=\mathcal{T}_{+}^{ab}-\mathcal{T}_{-}^{ab},
\label{explicit junction ij}
\end{equation}
where $\mathcal{T}_{\pm}^{ab}$ is the energy-momentum tensor given in Eq. (\ref{Hps}) associated
with the two sides of the shell.

\section{Thin-shell Wormhole Construction\label{Thin}}

To construct a thin-shell wormhole in third order Lovelock gravity, we use
the well-known cut-and-pase technique. Taking two copies of
asymptotically flat solutions of Lovelock gravity given by Eqs. (\ref{met})
and (\ref{Fgen}) and removing from each manifold the seven-dimensional region
described by
\begin{equation}
\Omega _{\pm }=\left\{ r_{\pm }\leq a; \ \ \ a>r_{h}\right\}
\end{equation}
we are left with two geodesically incomplete manifolds with the following
timelike hypersurface as boundaries
\begin{equation}
\Sigma _{\pm }=\left\{ r_{\pm }=a; \ \ \ a>r_{h}\right\} .
\end{equation}
Now identifying these two boundaries, $\Sigma _{+}=\Sigma _{-}=\Sigma $, we
leave with a geodesically complete manifold containing the two
asymptotically flat regions $\Omega _{+}$ and $\Omega _{-}$ which are
connected by a wormhole. The throat of the wormhole located at $\Sigma $
with metric
\begin{equation*}
ds_{\Sigma }^{2}=-d\tau ^{2}+a^{2}(\tau )d\Omega _{5}^{2},
\end{equation*}
where $\tau $\ is the proper time along the hypersurface $\Sigma $ and $%
a(\tau )$ is the radius of the throat. All the matter is concentrated on $%
\Sigma $.

To analyze such a thin-shell configuration, we need to use the modified
junction condition introduced in Sec. \ref{Junc}. Denoting the coordinates on $%
\Sigma $ by $\xi ^{a}=(\tau ,\theta ^{i};$ $i=1...5)$, the extrinsic
curvature associated with the two sides of the shell are
\begin{equation}
{\mathcal{K}}_{ab}^{\pm }=-n_{\rho }^{\pm }\left( \frac{\partial ^{2}X^{\rho
}}{\partial \xi ^{a}\partial \xi ^{b}}+\Gamma _{\mu \nu }^{\rho }\frac{%
\partial X^{\mu }}{\partial \xi ^{a}}\frac{\partial X^{\nu }}{\partial \xi
^{b}}\right) _{r=a},
\end{equation}
where $n_{\rho }^{\pm }$ are the units normal ($n_{\rho }n^{\rho }=1$) to
the surface $\Sigma $ in $\mathcal{M}$:
\begin{equation}
n_{\gamma }^{\pm }=\pm \left| g^{\mu \nu }\frac{\partial \mathcal{G}}{%
\partial X^{\mu }}\frac{\partial \mathcal{G}}{\partial X^{\nu }}\right|
\frac{\partial \mathcal{G}}{\partial X^{\gamma }}
\end{equation}
and $\mathcal{G}(r,\tau )$ is the equation of the boundary $\Sigma $:
\begin{equation}
\mathcal{G}(r,\tau )=r-a(\tau )=0.
\end{equation}
Using an orthonormal basis $\{e_{\hat{\tau}},e_{\hat{\imath}};$ $i=1...5\}$,
the components of extrinsic curvature tensor may be calculated as
\begin{equation}
\mathcal{K}_{~\hat{\tau}}^{\hat{\tau}} =\frac{\Gamma }{\Delta },  \hspace{1cm}
\mathcal{K}_{\hat{\jmath}}^{\hat{\imath}} =\frac{\Delta }{a}~\delta _{\hat{%
\jmath}}^{\hat{\imath}},  \label{Ki}
\end{equation}
where
\begin{equation*}
\Gamma =\ddot{a}+\frac{f^{\prime }\left( a\right) }{2},\text{ \ \ \ }\Delta =%
\sqrt{\dot{a}^{2}+f\left( a\right) ,}
\end{equation*}
and prime and overdot denote the derivative with respect to $a$ and $\tau $,
respectively. Equation (\ref{Ki}) shows that the form of the stress-energy
tensor on the shell is ${S}_{~\hat{b}}^{\hat{a}}=~\mbox{diag}~(-\sigma ,p~\delta
_{\hat{\jmath}}^{\hat{\imath}})$, where $\sigma $ is the surface energy
density and $p$ is the transverse pressure.

Now using the junction condition (\ref{explicit junction ij}), the
components of energy momentum tensor on the shell may be written as

\begin{eqnarray}
\sigma &=&-S_{\hat{\tau} }^{\hat{\tau} }=-\frac{\Delta }{4\pi a}\left\{ 5+\frac{40\alpha
_{2}}{a^{2}}[3(1+\dot{a}^{2})-\Delta ^{2}]\ +\frac{24\alpha _{3}}{a^{4}}[%
15(1+\dot{a}^{2})^{2}-10\Delta ^{2}(1+\dot{a}^{2})+3\Delta
^{4}]\right\} ,
\nonumber \\
p &=&S_{\hat{i}}^{\hat{i}}=\frac{1}{8\pi }\Bigg\{\frac{2\Gamma }{\Delta }+\frac{8\Delta }{%
a}+\frac{16\alpha _{2}}{a^{2}}\left[\frac{3\Gamma }{\Delta
}\left( 1+\dot{a}^{2}-\Delta^2\right) +\frac{2\Delta }{a}%
\left[3\left( 1+\dot{a}^{2}\right)-\Delta^2\right]+6\ddot{a}\Delta \right]  \nonumber \\
&&+\frac{48\alpha_{3}}{a^{4}}\left[ \frac{3\Gamma }{\Delta }\left( 1+\dot{a}^{2}-\Delta ^{2}\right)^2 +4\ddot{a}\Delta \left[ 3(1+\dot{a}^{2})-\Delta
^{2}\right] \right] \Bigg\}.  \label{pre}
\end{eqnarray}
One may note that the surface energy density and transverse pressure satisfy
the energy conservation equation:

\begin{equation}
\frac{d}{d\tau }\left( \sigma a^{5}\right) +p\frac{d}{d\tau }\left(
a^{5}\right) =0.  \label{eqco}
\end{equation}
The first term in Eq. (\ref{eqco}) represents the internal energy change of
the throat and the second term shows the work done by the throat's internal forces.

\section{Exoticity of the Matter on Shell\label{Exot}}

In this section, we consider the issue of energy condition on the shell for
the case of static configurations with $a=a_{0}$ and $\dot{a}=\ddot{a}=0$.
In our case the weak energy condition is satisfied provided $\sigma \geq 0$, and $%
\sigma +p\geq 0$. Using Eqs. (\ref{pre})\ one obtains
\begin{eqnarray}
\sigma _{0} &=&-\frac{1}{8\pi a_{0}\sqrt{f_{0}}}\left\{ 10f_{0}+\frac{%
80\alpha _{2}}{a_{0}^{2}}f_{0}(3-f_{0})+\frac{48\alpha _{3}}{a_{0}^{4}}%
f_{0}(15-10f_{0}+3f_{0}{}^{2})\right\} , \\
\sigma _{0}+p_{0} &=&\frac{1}{8\pi a_{0}\sqrt{f_{0}}}\Bigg\{%
(-2f_{0}+a_{0}f_{0}^{\prime })+\frac{8\alpha _{2}}{a_{0}^{2}}%
\{3a_{0}f_{0}^{\prime }(1-f_{0}) \\
&&-6f_{0}(3-f_{0})\}+\frac{24\alpha
_{3}}{a_{0}^{4}}\{3a_{0}f_{0}^{\prime
}(1-2f_{0}+f_{0}^{2})-2f_{0}(15-10f_{0}+3f_{0}{}^{2})\}\Bigg\},
\label{po}
\end{eqnarray}
where $f_{0}=f(a_{0})$ and $f_{0}^{\prime }=f^{\prime }(a_{0})$. In contrast
to the case of an Einsteinian thin-shell wormhole for which $\sigma <0$ and
therefore the weak energy condition is violated \cite{Lem}, here we can have
thin-shell wormholes with normal matter on shell. In order to investigate
the exoticity of the matter, we calculate the amount of matter on shell which is
\begin{equation}
\digamma =\int drd\Omega _{5}[\sigma _{0}\delta (r-a_{0})+p_{r}].
\end{equation}
For our case, the shell does not exert radial pressure, $p_{r}=0$, and
therefore the amount of matter on shell is
\begin{equation}
\digamma =\pi ^{3}a_{0}^{5}\sigma _{0}=\pi ^{2}\sqrt{f_{0}}\left\{ -\frac{5}{%
4}a_{0}^{4}+10\alpha _{2}a_{0}^{2}(-3+f_{0})+6\alpha _{3}\left(
-15+10f_{0}-3f_{0}^{2}\right) \right\}.  \label{FF}
\end{equation}
In Einstein gravity, $%
\alpha _{2}=\alpha _{3}=0$, the matter is exotic both for Reissner-Nordstrom
$(Q\neq 0)$ and Schwarzschild $(Q=0)$ thin-shell wormholes as one can see
from Eq. (\ref{FF}). In Gauss-Bonnet gravity ($\alpha _{3}=0$) with $\alpha
_{2}<0$, one may have thin-shell wormholes supported by normal matter
\cite{Mazha}.

\begin{figure}[h]
\centering
{\includegraphics[width=.3\textwidth]{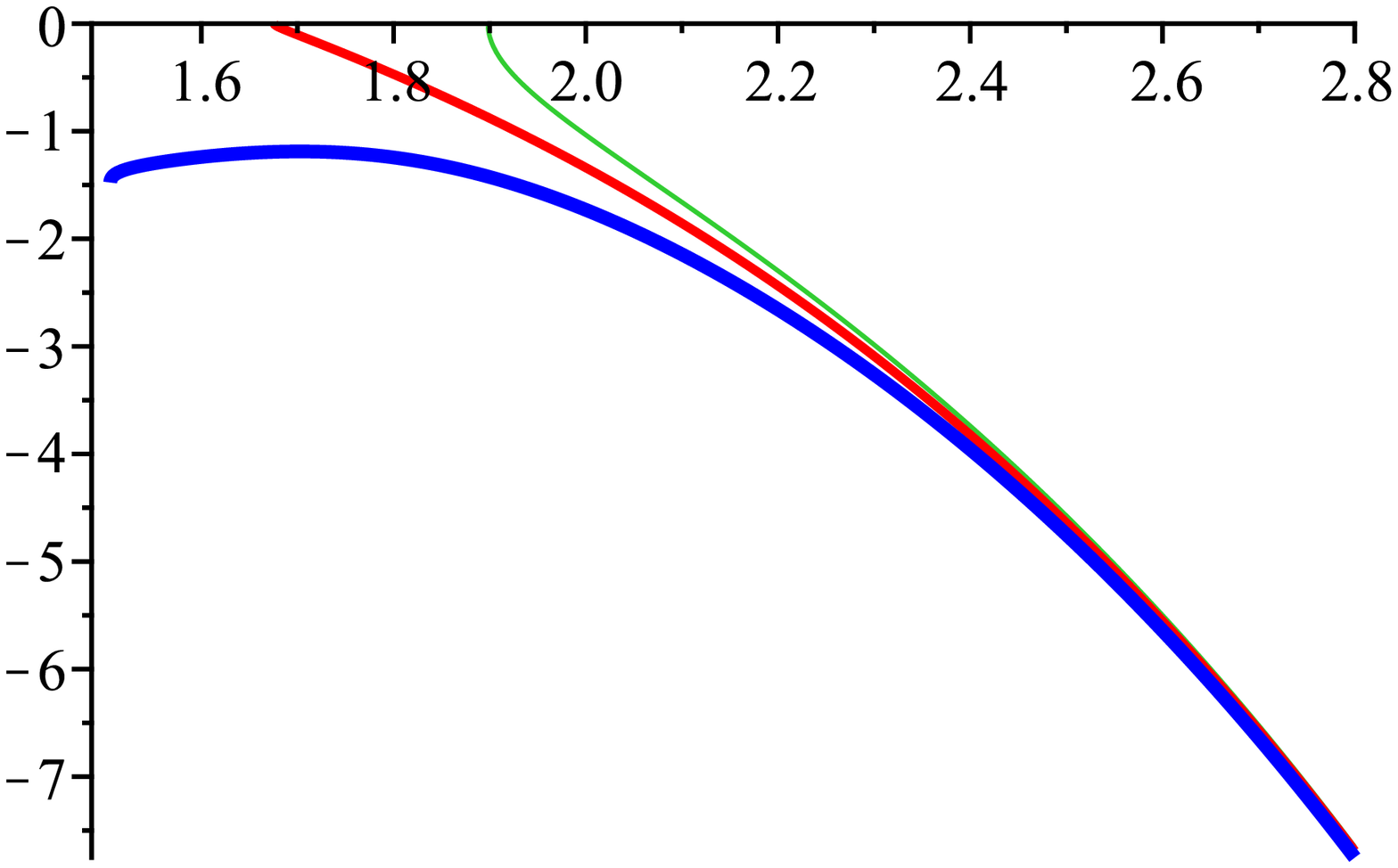}\qquad} {%
\includegraphics[width=.3\textwidth]{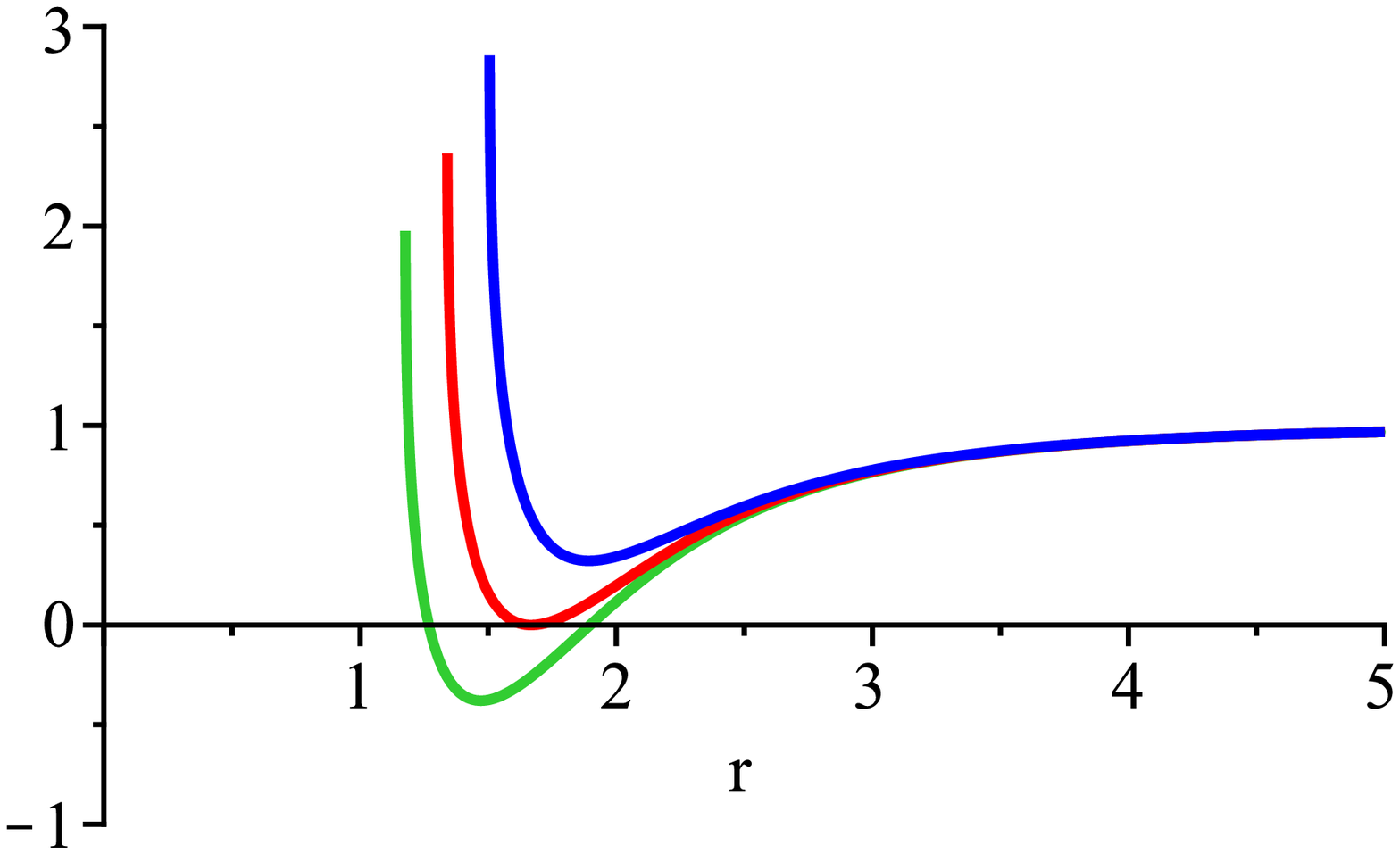}}
\caption{Right: $f(r)$ versus $r$; Left: $\digamma/100$ versus $a_{0}$ for $\alpha=1$, $m=20$,
$q>q_{ext}$, $q=q_{ext}$ and $q<q_{ext}$ from up to down,
for the right figure and down to up for the left figure, respectively.} \label{f12}
\end{figure}

Now, we investigate the condition that thin-shell wormhole may be supported
by normal matter in third order Lovelock gravity. For the special case $%
\alpha _{3}=2{\alpha _{2}}^{2}=\alpha ^{2}/72$ , Eq. (\ref{FF}) shows that $%
\digamma <0$, and therefore the on shell matter is exotic. But, one can
choose the parameters of the metric function such that the amount of exotic
matter on the throat to be as less as possible. For instance, as one can see in
Fig. \ref{f12}, the amount of exotic matter decreases as $q$ increases.

\begin{figure}[h]
\centering
{\includegraphics[width=.3\textwidth]{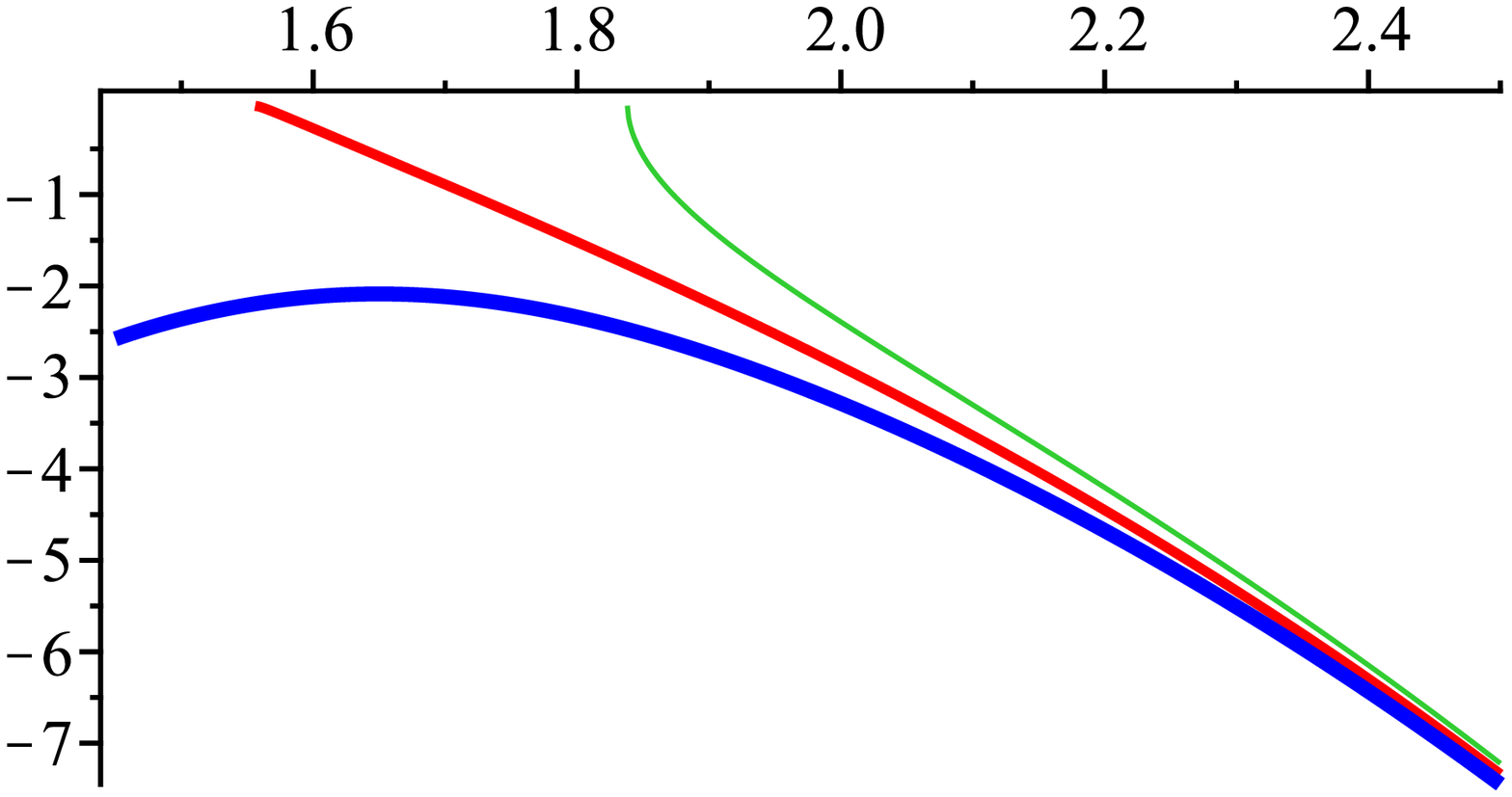}\qquad} {%
\includegraphics[width=.3\textwidth]{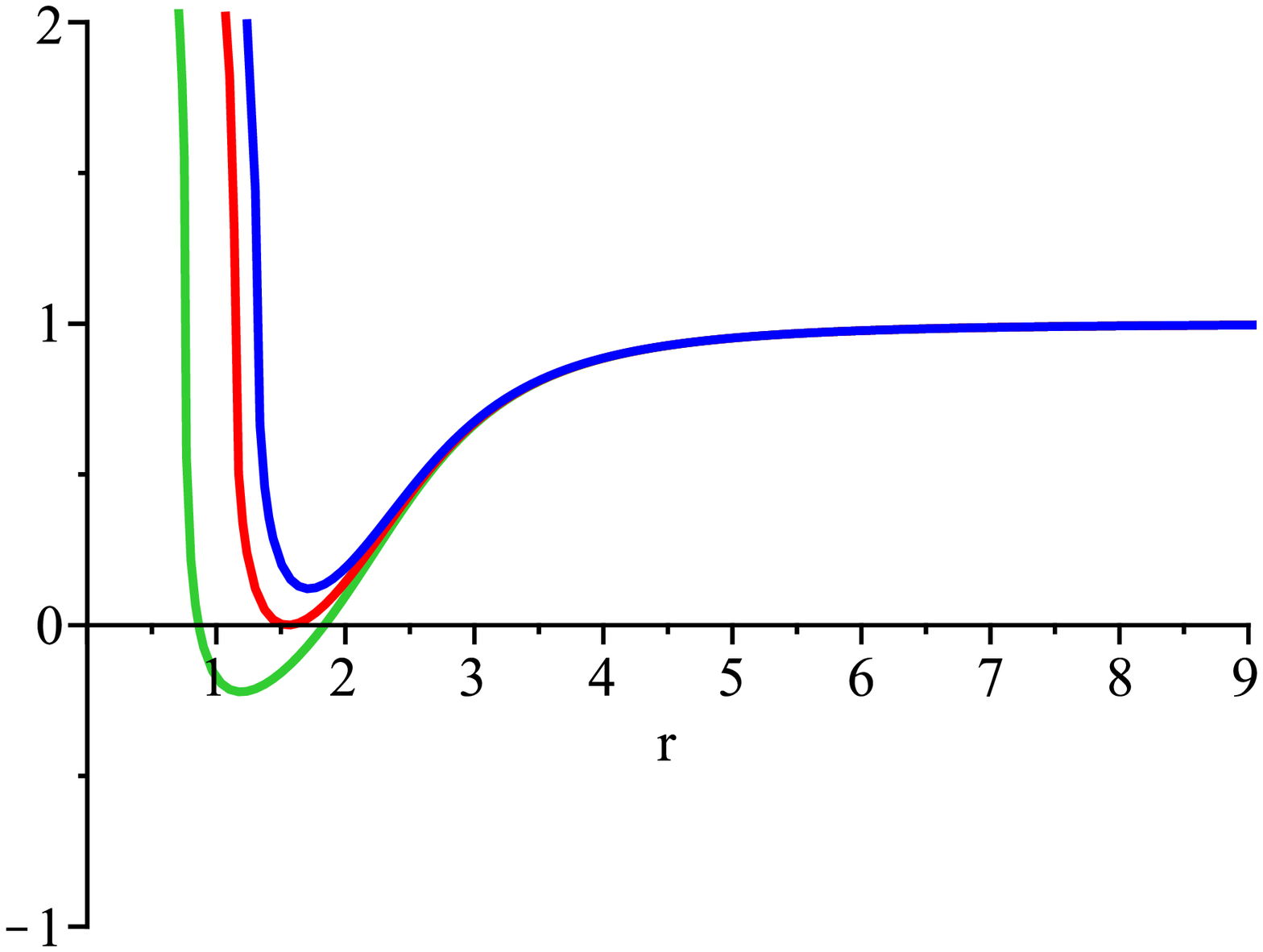}}
\caption{Right: $f(r)$ versus $r$; Left: $\digamma/100$ versus $a_{0}$ for $\alpha_{2}=0.2$,
$\alpha_{3}=0.4$, $m=30$, $q>q_{ext}$, $q=q_{ext}$ and
$q<q_{ext}$ from up to down, for the right figure and down to up for the left figure, respectively.} \label{f34}
\end{figure}

\begin{figure}[h]
\centering
{\includegraphics[width=.3\textwidth]{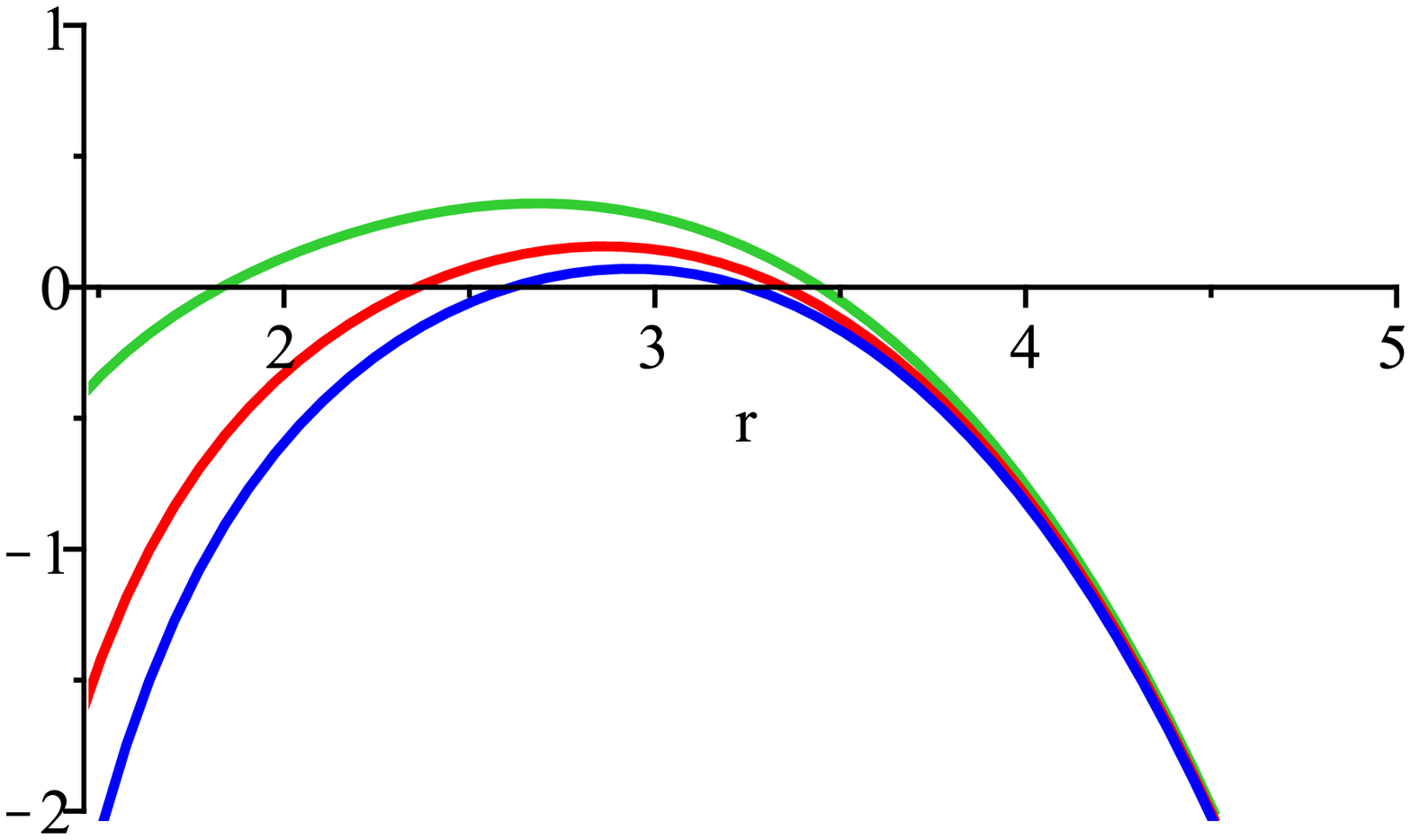}\qquad} {%
\includegraphics[width=.3\textwidth]{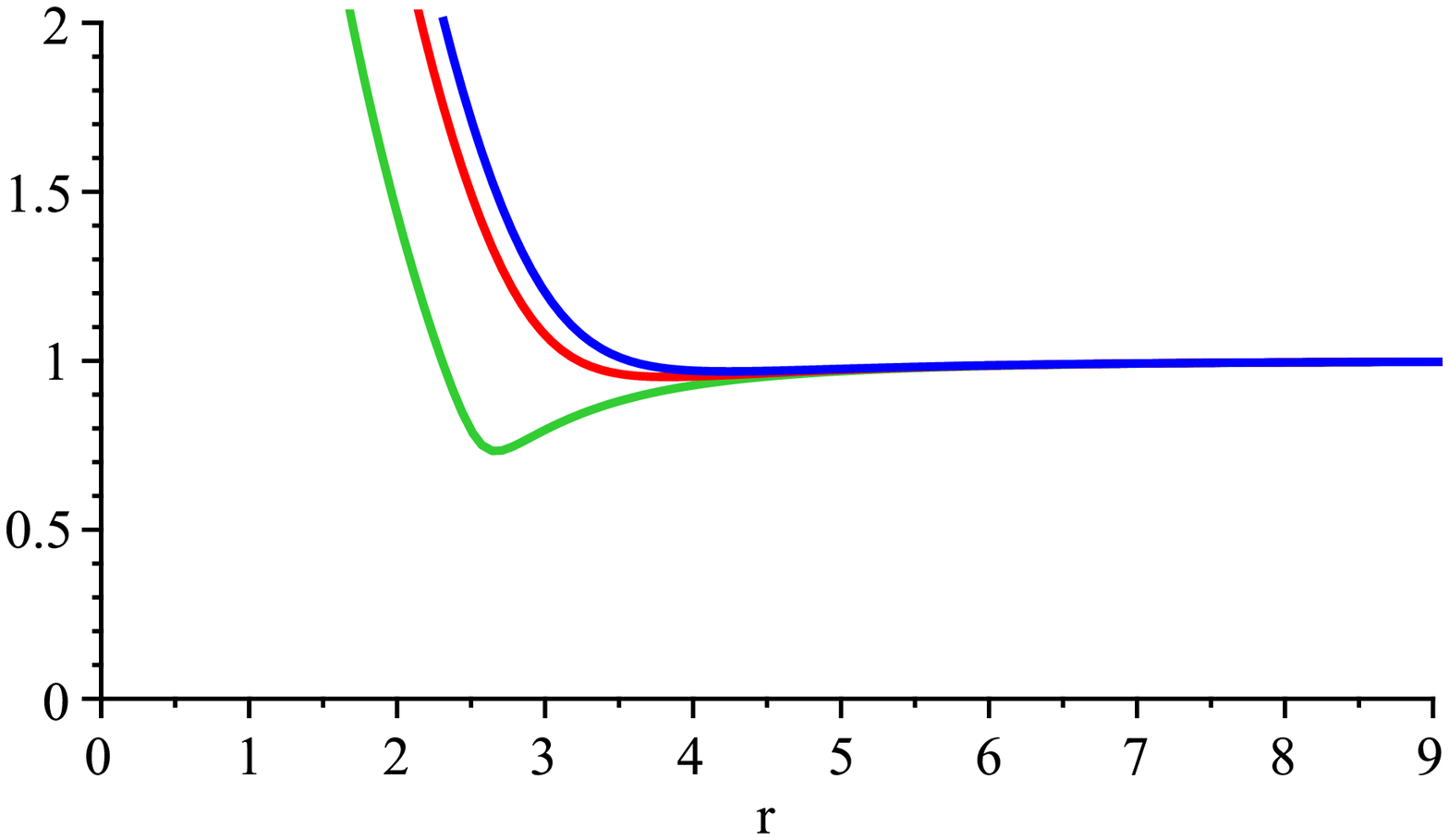}}
\caption{Right: $f(r)$ versus $r$; Left: $\digamma/1000$ versus
$a_{0}$ for $\alpha_{2}=-0.8 $, $\alpha_{3}=0.5$, $m=20$,
$q=57$, 46.5 and 23.72 from up to down, for the right figure and down to up for the left figure, respectively.} \label{f65}
\end{figure}

\begin{figure}[h]
\centering {\includegraphics[width=.4\textwidth]{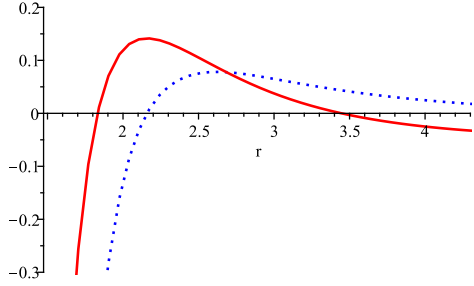}}
\caption{$\protect\sigma_{0}$ (line) and $\protect\sigma_{0}
+p_{0}$ (dotted) versus $r$ for $\alpha_{2}=-0.8$, $\alpha_{3}=0.5$,
$m=20$ and $q=23.72$.} \label{wloop}
\end{figure}

For the general solutions of third order Lovelock gravity with positive
values of $\alpha _{2}$ and $\alpha _{3}$, the matter on the shell is
exotic as one can see in Fig. \ref{f34}. But, for $\alpha _{3}>0$ and $%
\alpha _{2}<0$, $\digamma $ can be positive and therefore the matter may be
normal, as one can see in Fig \ref{f65}. For this case as Fig \ref{wloop} shows, there exists a region with $%
\sigma _{0}\geq 0$ and $\sigma _{0}+p_{0}\geq 0$ and therefore
WEC is are satisfied. Since $0<f_{0}<1$, the factor of $\alpha _{3}$ in Eq. (%
\ref{FF}), $-15+10f_{0}-3f_{0}^{2}$, is negative and therefore the
amount of normal matter for negative $\alpha _{2}$ decreases as
$\alpha _{3}$ increases. Also, in this case the amount of normal
matter decreases as the charge $q$ increases.

\section{STABILITY ANALYSIS \label{Stab}}

In this section, we perform a stability analysis under a linear perturbation
such that the spherical symmetry of the wormhole configuration is preserved.
To analyze the stability, we use a cold equation of state $%
p=p(\sigma )$ with $\eta =dp/d\sigma $. We consider a small radial
perturbation around a static solution with radius $a_{0}$. In this
case, one may write $p\simeq p_0+\eta_0(\sigma-\sigma_0)$, where
$p_{0}$, $\sigma _{0}$ and $\eta_0$ are the transverse pressure,
surface energy density and $(dp/d\sigma)$ at $a=a_{0}$,
respectively. Using this linear equation of state and Eq.
(\ref{eqco}), one obtains
\begin{equation}
\sigma \left( a\right) =\left( \frac{\sigma _{0+}p_{0}}{1+\eta_0 }\right)
\left( \frac{a_{0}}{a}\right) ^{5\left( 1+\eta_0 \right) }+\frac{\eta_0 \sigma
_{0-}p_{0}}{1+\eta_0 }.  \label{EnD}
\end{equation}
Now Eq. (\ref{pre}), which is the equation of motion for the radius of the throat, can be written as
\begin{eqnarray}
&&5a^{4}+40\alpha _{2}a^{2}\left[ 2(1+\dot{a}^{2})+1-{f(a)}\right]
+24\alpha
_{3}\Big\{8(1+\dot{a}^{2})^{2}  \nonumber \\
&&+4[1-f(a)](1+\dot{a}^{2})+3[1-f(a)]^{2}]\Big\}=-\frac{4\pi a^{5}\sigma (a)}{%
\sqrt{\dot{a}^{2}+f(a)}},  \label{eqgen}
\end{eqnarray}
where $\sigma (a)$ is given in Eq. (\ref{EnD}).

\begin{figure}[h]
\centering
{\includegraphics[width=.34\textwidth]{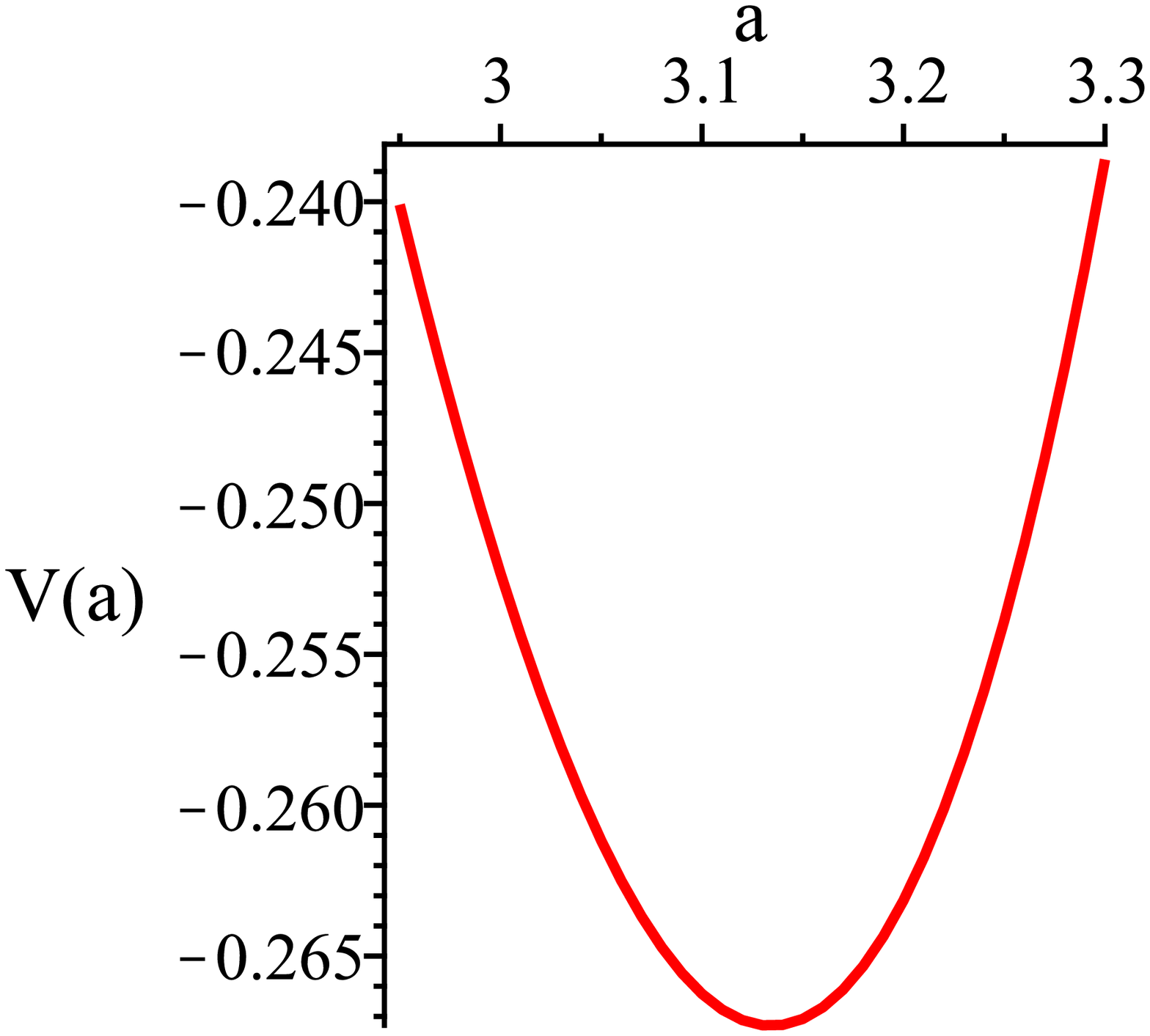}\qquad} {%
\includegraphics[width=.3\textwidth]{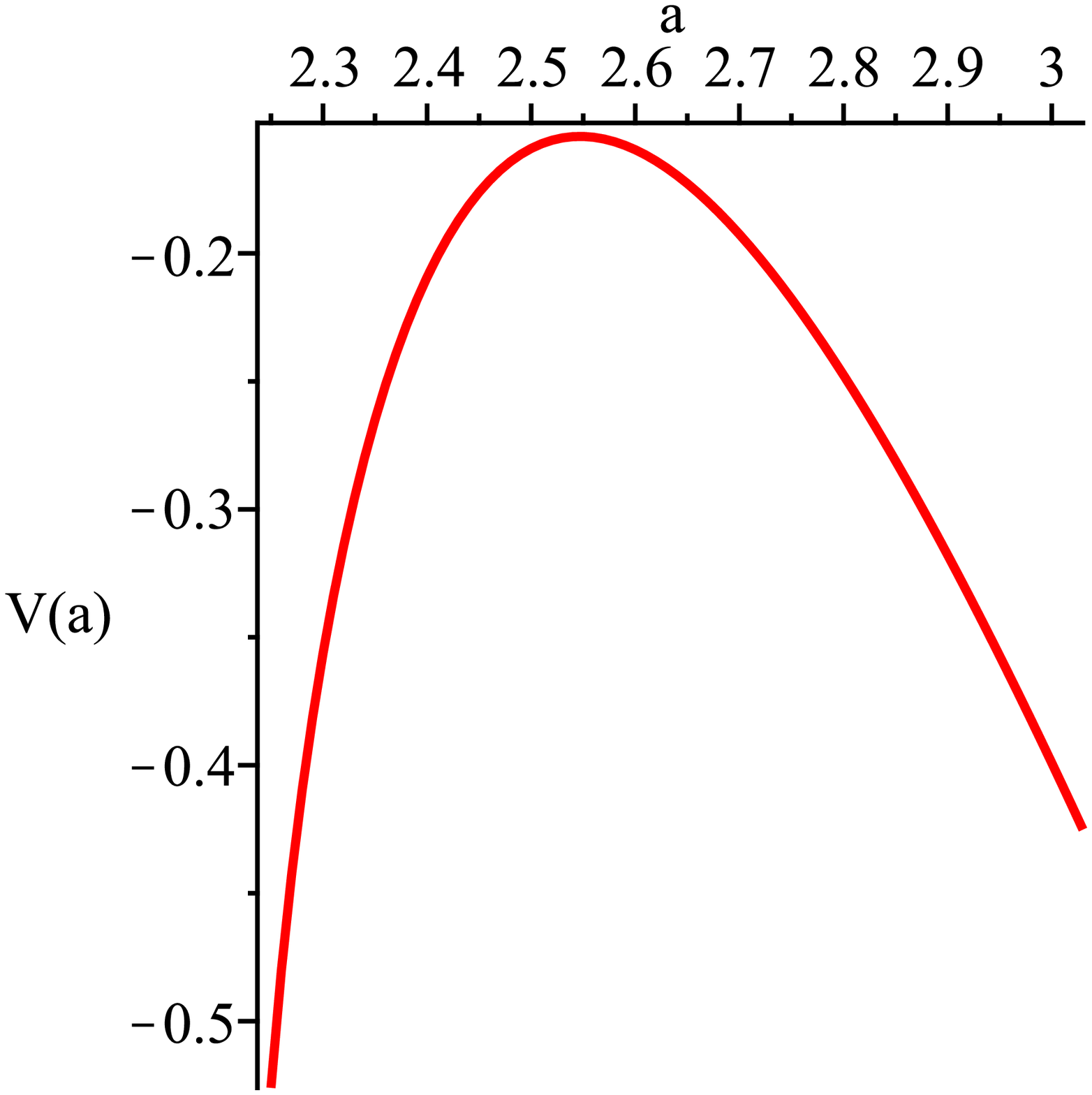}}
\caption{$V(a)$ versus $a$ for $\alpha_{2}=-0.8$,
$\alpha_{3}=0.5$, $m=20$, $q=23.72$, and $\eta=1.2$ (right); $\eta
=-1.8$ (left).} \label{Stab1}
\end{figure}

\begin{figure}[h]
\centering
{\includegraphics[width=.3\textwidth]{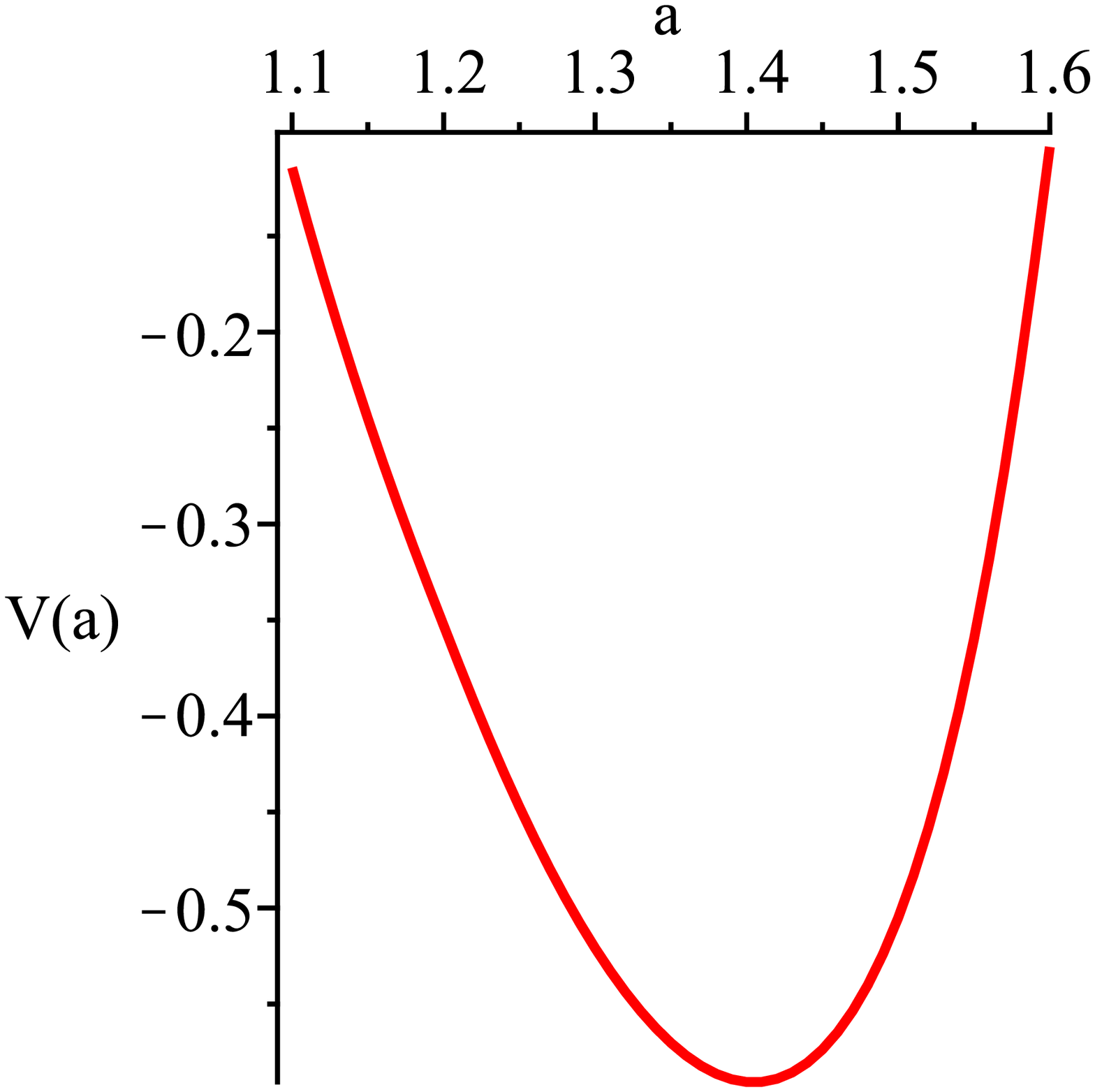}\qquad} {%
\includegraphics[width=.3\textwidth]{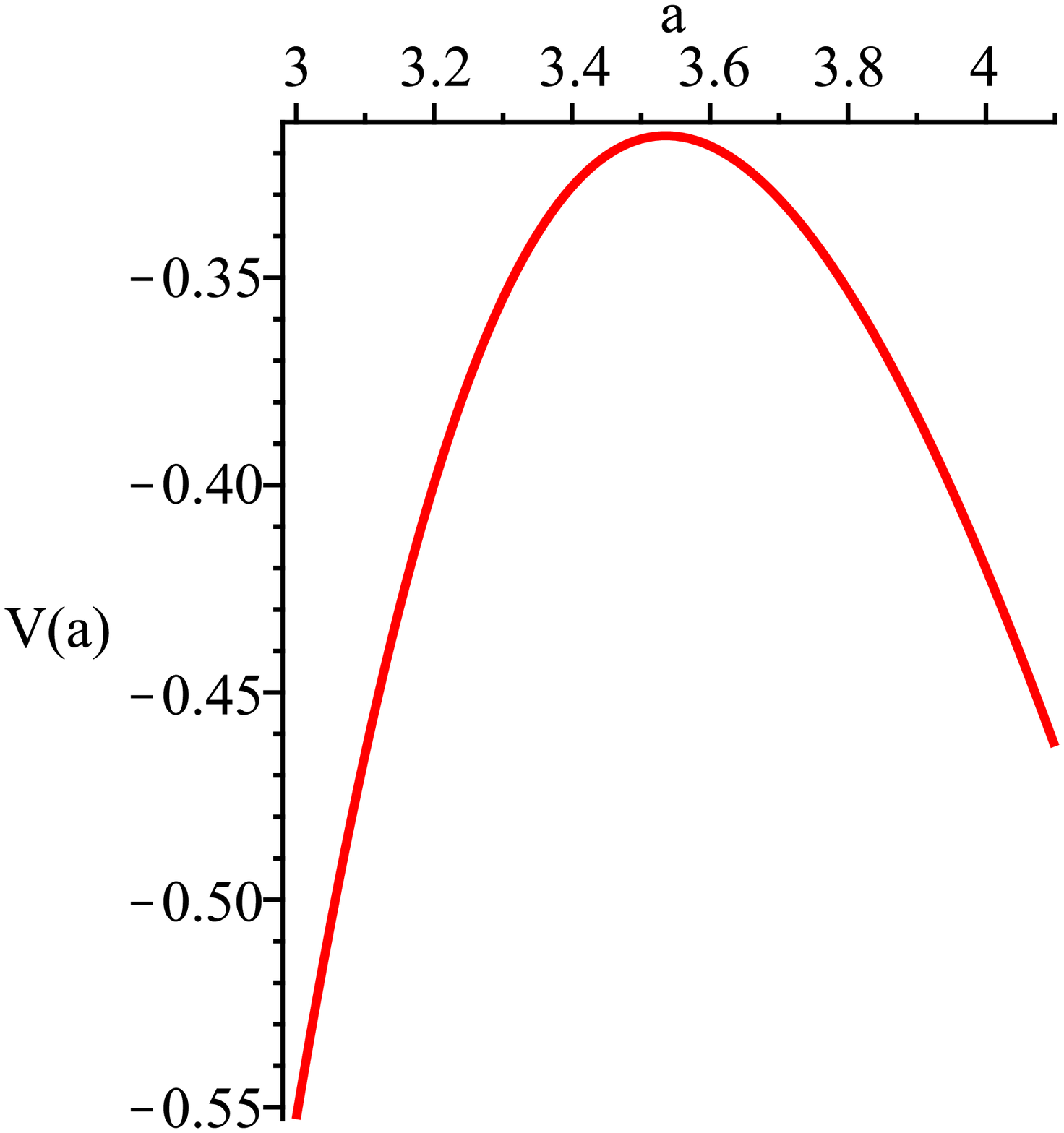}}
\caption{$V(a)$ versus $a$ for $\alpha_{2}=0.2$, $\alpha_{3}=0.4$,
$m=30$, $q=7.9$ and $\eta=0.9$ (right); $\eta =-1.8$ (left).}
\label{Stab2}
\end{figure}

In principle, one may solve Eq. (\ref{eqgen}) for $\dot{a}^{2}$
and obtain the potential $V(a)$ in the equation
$\dot{a}^{2}=-V(a)$. Then, the wormhole with radius $a_{0}$ is
linearly stable provided the potential $V(a)$ is negative and minimum at
$a=a_{0}$. In third order Lovelock gravity, one encounters with a
fifth order algebraic equation for $\dot{a}^{2}$ and therefore one
may perform stability analysis numerically. Numerical calculations
are shown in Figs. \ref{Stab1} and \ref{Stab2}. As these figures
show, the wormholes are stable provided the derivative of surface pressure
density with respect to surface energy density at the throat, $\eta_0$, is negative and
the throat radius $a_{0}$ is chosen suitable.

\section{CLOSING REMARKS}
In this paper, we first use the
well-known cut-and-pase technique, and constructed the
asymptotically flat thin-shell wormholes of Lovelock gravity in
seven dimensions. We calculated the components of energy momentum
tensor on shell through the use of the general junction condition.
We found that the matter on the throat is exotic if both $\alpha
_{2}$ and $\alpha _{3}$ are positive. However, the amount of
exotic matter on shell reduces as the charge of the wormhole
increases. In the case of negative $\alpha_2$ and positive
$\alpha_3$, one may have a region for the throat radius with
$\sigma _{0}\geq 0$ and $\sigma _{0}+p_{0}\geq 0$, and therefore
WEC is satisfied. That is, one may have wormholes
with normal matter provided $\alpha _{2}<0$ and $\alpha _{3}>0$.
In this case, the amount of normal matter decreases as the third
order Lovelock parameter increases. Finally, we applied a linear
stability analysis against symmetry preserving perturbation and
found that the wormholes with suitable throat radius are stable provided $\eta_0=(dp/d\sigma)_{a_0}<0$.

\acknowledgements This work was
supported by the Research Institute for Astrophysics and Astronomy
of Maragha.


\begin{thebibliography}{99}
\bibitem{MT}  M. S. Morris and K. S. Thorne, Am. J. Phys. \textbf{56}, 395
(1986).

\bibitem{MTY}  M. S. Morris, K. S. Thorne, and U. Yurtsever, Phys. Rev.
Lett. \textbf{61}, 1446 (1988); M. Visser, Lorentzian Wormholes: From
Einstein to Hawking (American Institute of Physics, New York, 1995).

\bibitem{viskardad}  M. Visser, S. Kar and N. Dadhich, Phys. Rev. Lett.
\textbf{90}, 201102 (2003).

\bibitem{Mart Rich}  M. G. Richarte, Phys. Rev. D \textbf{82}, 044021
(2010).

\bibitem{Deh1}  M. H. Dehghani and Z. Dayyani, Phys. Rev. D \textbf{79},
064010 (2009).

\bibitem{Vis}  M. Visser, Phys. Rev. D \textbf{39,} 3182 (1989); Nucl. Phys.
B \textbf{328}, 203 (1989).

\bibitem{VisP}  E. Poisson and M. Visser, Phys. Rev. D \textbf{52}, 7318 (1995).

\bibitem{Thin}  S. W. Kim, Phys. Lett. A \textbf{166}, 13 (1992); F. S. N.
Lobo, Class. Quant. Grav. \textbf{21}, 4811 (2004); J. P. S. Lemos and F. S.
N. Lobo, Phys. Rev. D \textbf{69}, 104007 (2004); E. F. Eiroa and C.
Simeone, \emph{ibid.} \textbf{70}, 044008 (2004); E. F. Eiroa and C.
Simeone, \emph{ibid.} \textbf{71}, 127501 (2005); F. Rahaman, M. Kalam, and
S. Chakraborty, Gen. Relativ. Gravit. \textbf{38}, 1687 (2006); C. Bejarano, E. F.
Eiroa, and C. Simeone, Phys. Rev. D \textbf{75}, 027501 (2007); F. Rahaman,
M. Kalam, and S. Chakraborty, Int. J. Mod. Phys. D \textbf{16}, 1669 (2007);
F. Rahaman, M. Kalam, K. A. Rahman, and S. Chakraborty, Gen. Relativ. Gravit.
\textbf{39}, 945 (2007); E. Gravanis and S. Willison, Phys. Rev. D \textbf{75}%
, 084025 (2007); M. G. Richarte and C. Simeone, Int. J. Mod. Phys. D \textbf{17},
1179 (2008).

\bibitem{Lobo}  F. S. N. Lobo and P. Crawford, Class. Quant. Grav. \textbf{21%
}, 391 (2004).

\bibitem{Lem}  F. Rahaman, M. Kalam, and S. Chakraborty, Gen. Relativ.
Gravit. \textbf{38}, 1687 (2006); J. P. S. Lemos and F. S. N. Lobo, Phys.
Rev D \textbf{78}, 044030 (2008); G. A. S. Dias and J. P. S. Lemos,
arXiv:1008.3376.

\bibitem{Axion}  E. F. Eiroa, Phys. Rev. D \textbf{78}, 024018 (2008); A. A.
Usmani, F. Rahaman, S. Ray, Sk. A. Rakib, and Z. Hasan, Gen. Relativ.
Gravit. \textbf{42}, 2901 (2010); P. K.
F. Kuhfittig, Acta Phys. Polonica B,\textbf{41}, 2017
(2010); E. F. Eiroa and C. Simeone, Phys.
Rev. D \textbf{81}, 084022 (2010).

\bibitem{Dil}  C. Bejarano and E. F. Eiroa, Phys. Rev. D \textbf{84}, 064043 (2011).

\bibitem{Brans}  X. Yue and S. Gao, Phys. Lett. A \textbf{375}, 2193 (2011);
Francisco S. N. Lobo, Miguel A. Oliveira, Phys. Rev. D \textbf{81}, 067501 (2010).

\bibitem{Gauss}  E. Gravanis and S. Willison, Phys. Rev. D \textbf{75},
084025 (2007); G. Dotti, J. Oliva, and R. Troncoso, \emph{ibid}. \textbf{76}, 064038 (2007);
H. Maeda and M. Nozawa, \emph{ibid}. \textbf{78}, 024005
(2008); M. Richarte and C. Simeone, \emph{ibid}. \textbf{76}, 087502 (2007);
\emph{Erratum-ibid.} D \textbf{77}, 089903 (2008); M. Thibeault, C. Simeone,
and E. F. Eiroa, Gen. Relativ. Gravit. \textbf{38}, 1593 (2006).

\bibitem{Deh2}  M. H. Dehghani and R. Pourhasan, Phys. Rev. D \textbf{79},
064015 (2009); M.H. Dehghani and R.B. Mann, J. High Energy Phys. \textbf{07}%
, 019 (2010); M. H. Dehghani and Sh. Asnafi, Phys. Rev. D \textbf{84},
064038 (2011).

\bibitem{Lg3}  X.H. Ge, S.J. Sin, S.F. Wu and G.H. Yang, Phys. Rev. D
\textbf{80}, 104019 (2009); J. de Boer, M. Kulaxizi and A. Parnachev, J.
High Energy Phys. \textbf{06}, 008 (2010); X.O. Camanho and J. D. Edelstein,
[arXiv:0912.1944]; F. W. Shu,  Phys. Lett. B \textbf{685}, 325 (2010).

\bibitem{Lov}  D. Lovelock, J. Math. Phys. \textbf{12}, 498n (1971); D.
Lovelock, Aequationes Math. \textbf{4}, 127 (1970).

\bibitem{Isr}  W. Israel, Nuovo Cimento \textbf{44B}, 1 (1966); N. Sen, Ann.
Phys. (Leipzig) \textbf{73}, 365 (1924); K. Lanczos, \textit{ibid.} \textbf{%
74}, 518 (1924); G. Darmois, M\'{e}morial des Sciences Math\'{e}matiques,
Fascicule XXV ch. V (Gauthier-Villars, Paris, 1927).

\bibitem{Dav}  S. C. Davis, Phys. Rev. D 67, 024030 (2003).

\bibitem{Grav} E. Gravanis and S. Willison, J. Geom. Phys. \textbf{57}, 1861 (2007);
O. Miskovic and R. Olea, J. High Energy Phys. \textbf{10}, 28 (2007).


\bibitem{Deh2005}  M. H. Dehghani and M. Shamirzaie, Phys. Rev. D \textbf{72}, 124015
(2005).

\bibitem{DBS}C. Teitelboim and J. Zanelli, Class. and Quant.Grav. 4, L125 (1987);
M. H. Dehghani, N. Bostani, A. Sheykhi, Phys. Rev. D \textbf{73}, 104013 (2006).

\bibitem{Mazha} S. Habib Mazharimousavi, M. Halilsoy and Z. Amirabi, Phys. Rev. D \textbf{81}, 104002
(2010); Class. Quant. Grav. \textbf{28}, 025004 (2011).
\end{thebibliography}
\end{document}